\documentclass{aipproc}
\layoutstyle{8x11double}
\usepackage{psfrag}

\begin{document}

\title
      {Temporal properties of short and long gamma-ray bursts}



\author{S.\,McBreen}{
  address={Department of Experimental Physics, University College
Dublin, Dublin 4, Ireland},
  email={smcbreen@bermuda.ucd.ie},
}

\author{F.\,Quilligan}{
  address={Department of Experimental Physics, University College
Dublin, Dublin 4, Ireland},
  email={fquillig@bermuda.ucd.ie},
}

\author{B.\,McBreen}{
  address={Department of Experimental Physics, University College
Dublin, Dublin 4, Ireland},
  email={bmcbreen@ucd.ie},
}

\author{L.\,Hanlon}{
  address={Department of Experimental Physics, University College
Dublin, Dublin 4, Ireland},
  email={lhanlon@bermuda.ucd.ie},
}

\author{D.\,Watson}{
  address={X-Ray Astronomy Group, Department of Physics and Astronomy, Leicester University, Leicester LE1 7RH, UK},
  email={wat@star.le.ac.uk},
}

\copyrightyear  {2001}

\begin{abstract}
A temporal analysis was performed on a sample of 100 bright short GRBs with T$_{\rm 90}$$<$$2\,$s from the BATSE Current Catalog along with
a similar analysis on 319 long bright GRBs with T$_{\rm 90}$$>$$2\,$s from the same catalog.
The short GRBs were denoised using a median filter and the long GRBs were denoised using a wavelet method.
Both samples were subjected to an automated pulse selection algorithm to objectively determine the effects of
neighbouring pulses. The rise times, fall times, FWHM, pulse amplitudes and areas were measured and their
frequency distributions are presented. The time intervals between pulses were also measured. The
frequency distributions of the pulse properties were found to be similar and consistent with lognormal distributions for both the short and
long GRBs.  The time intervals between the pulses and the pulse amplitudes of neighbouring pulses were found to be correlated
with each other.  The same emission mechanism can account for the two sub-classes of GRBs.
\end{abstract}

\date{\today}

\maketitle

\section{Introduction}
It has been recognised that GRBs may occur in two sub-classes
based on spectral hardness and duration with T$_{\rm 90}$$>$$2\,$s and
T$_{\rm 90}$$<$$2\,$s \citep{kmf:1993,nsb:2000,paciesas:2001}.  The
bimodal distribution can be fit by two Gaussian distributions to
the logarithmic durations \citep{mhlm:1994}. A variety of
statistical methods have been applied to the temporal properties
of the long GRBs with T$_{\rm 90}$$>$$2\,$s. It is important to
compare the temporal properties of the long and short GRBs to
determine the similarities and differences between the two
classes in an objective way. Detailed temporal analyses have been
performed on a large sample of  short and long bright GRBs. The
results from the long sample \citep{quillig:2002} can be used as templates for
comparison with a similar analysis of short GRBs.

\section{Data Analysis}
The sample of 100 short GRBs was selected from the Time Tagged Event data at 5 ms from the BATSE Current Catalog.
The four energy channels were combined to maximise the signal to noise ratio.
The sample of 319 long GRBs was selected from the "discsc" 64ms data also from the BATSE Current Catalog.
The energy channels were combined as for the short GRBs.
Bursts from both samples were background subtracted by selecting a pre- and/or post-burst section.

A median filter was used to denoise the short GRBs
\citep{smcb:2001}
and a wavelet method
\citep{quillig:2002} was used 
 to denoise the long GRBs. The
same pulse selection method was applied to each sample of
denoised GRBs. The pulses selected had a threshold of 5 $\sigma$
above  background ($\mathrm{\tau}_{\rm \sigma}$ $\geq$ 5) and
were isolated from adjacent pulses by at least 50\%
($\mathrm{\tau}_{\rm i}$ $\geq$ 50\%).  A value of
$\mathrm{\tau}_{\rm i}$ $\geq$ 50\% implies that the two minima
on either side of the pulse maximum must be at or below half the
maximum value. A total of 313 pulses were selected from the
sample of short GRBs with $\mathrm{\tau}_{\rm \sigma}$ $\geq$ 5
and 181 of these had $\mathrm{\tau}_{\rm i}$ $\geq$ 50\%. A total
of 3358 pulses with $\mathrm{\tau}_{\rm \sigma}$ $\geq$ 5 were
selected from the sample of long GRBs, 1575 of which had
$\mathrm{\tau}_{\rm i}$ $\geq$ 50\%.

\section{Results}
The distributions of rise times ($t_{\rm r}$), fall times
($t_{\rm f}$) and full width at half maxima (FWHM) for the
isolated pulses  ($\mathrm{\tau}_{\rm i}$ $\geq$ 50\%) are
presented in Fig. 1. The distribution of time intervals between
the pulses ($\Delta$T) with $\mathrm{\tau}_{\rm \sigma}$ $\geq$ 5
is also given in Fig. 1. The distributions of pulse amplitudes
and areas are given in Figs. 2 and 3 for the isolated pulses
observed by two BATSE large area detectors. The median values of
the distributions are presented in Table 1. The Spearman Rank
Order correlation coefficients $\rho$ along with associated
probabilities for the time intervals separated by N pulses are
given in Table 2. The $\Delta$T values are normalised by T$_{\rm
90}$ for each burst and show that there is a high degree of
correlation over many intervals for the long GRBs.
The Spearman Rank Order correlation coefficents for the pulse amplitudes with N are listed in Table 3
for the short and long bursts.

\begin{figure}[hbp]
    \begin{minipage}{\columnwidth}
    \begin{center}
    \leavevmode
\vspace{0.5em}
        \psfrag{line 1}[b]{\tiny Long GRBs}
        \psfrag{line 2}[b]{\tiny Short GRBs}
        \psfrag{xlab}[b]{\small Rise Times (sec)}
        \psfrag{ylab}[b]{\small Frequency}
\includegraphics[height=5cm]{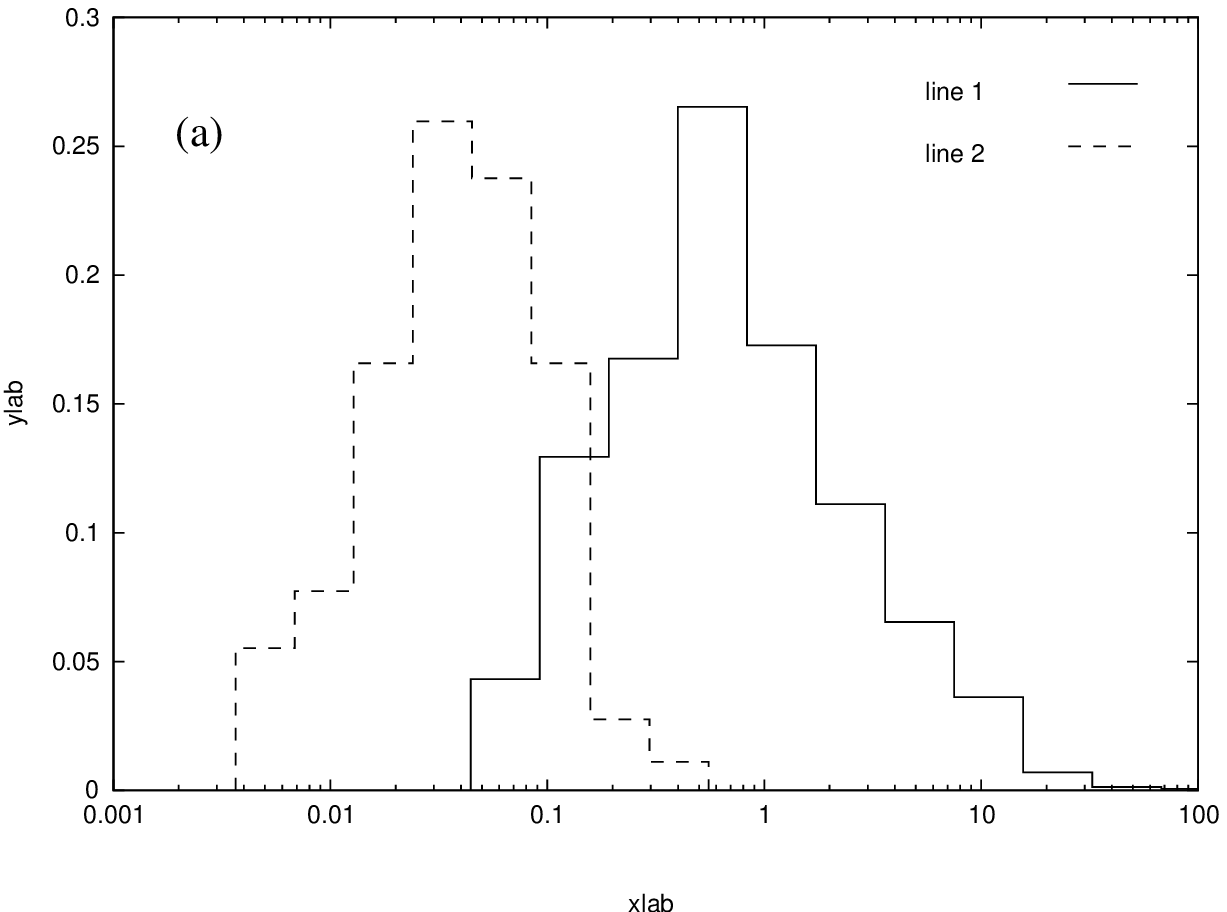} 
\vspace{0.5em}
        \psfrag{xlab}[b]{\small Fall Times (sec)}
        \psfrag{ylab}[b]{\small Frequency}
\includegraphics[height=5cm]{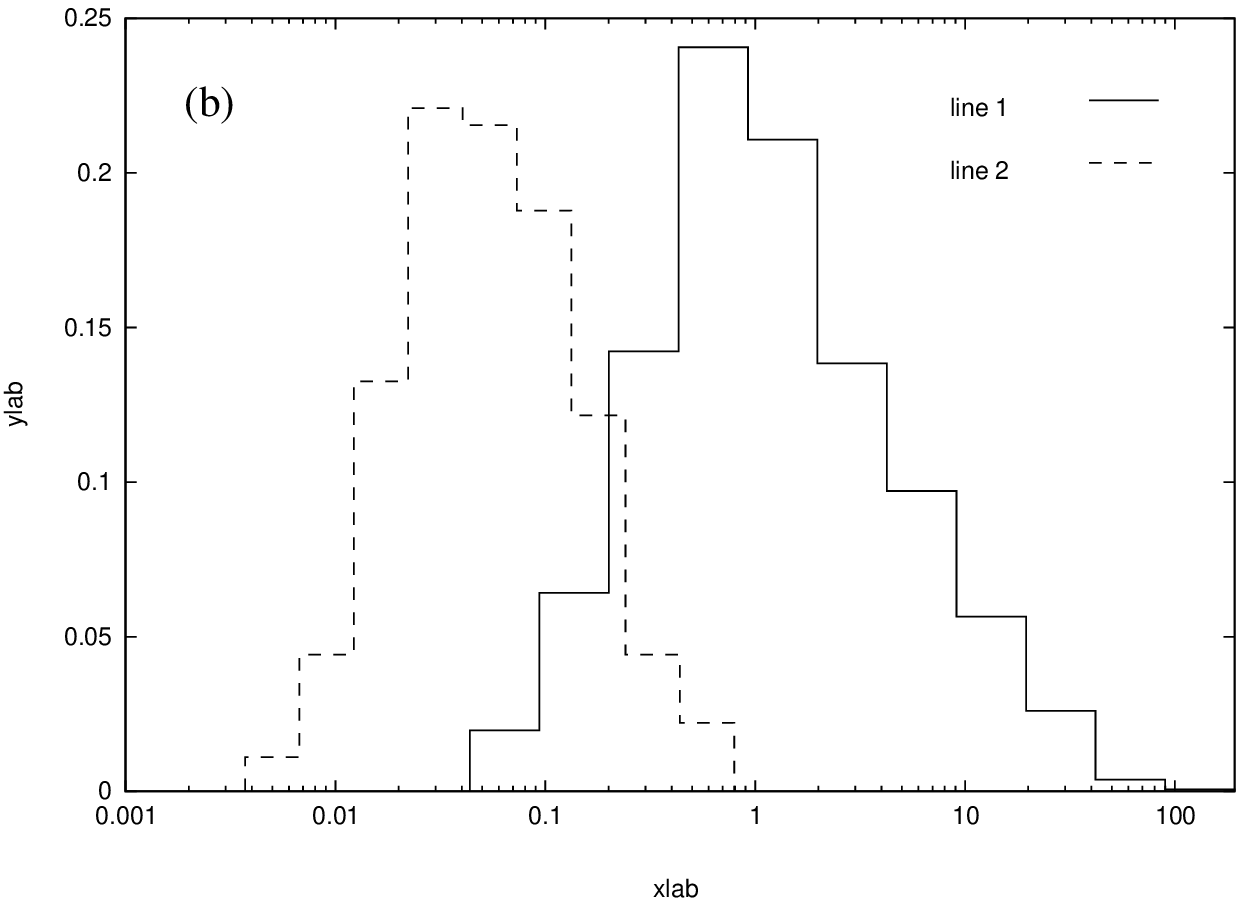} 
\vspace{0.5em}
        \psfrag{xlab}[b]{\small FWHM (sec)}
        \psfrag{ylab}[b]{\small Frequency}
\includegraphics[height=5cm]{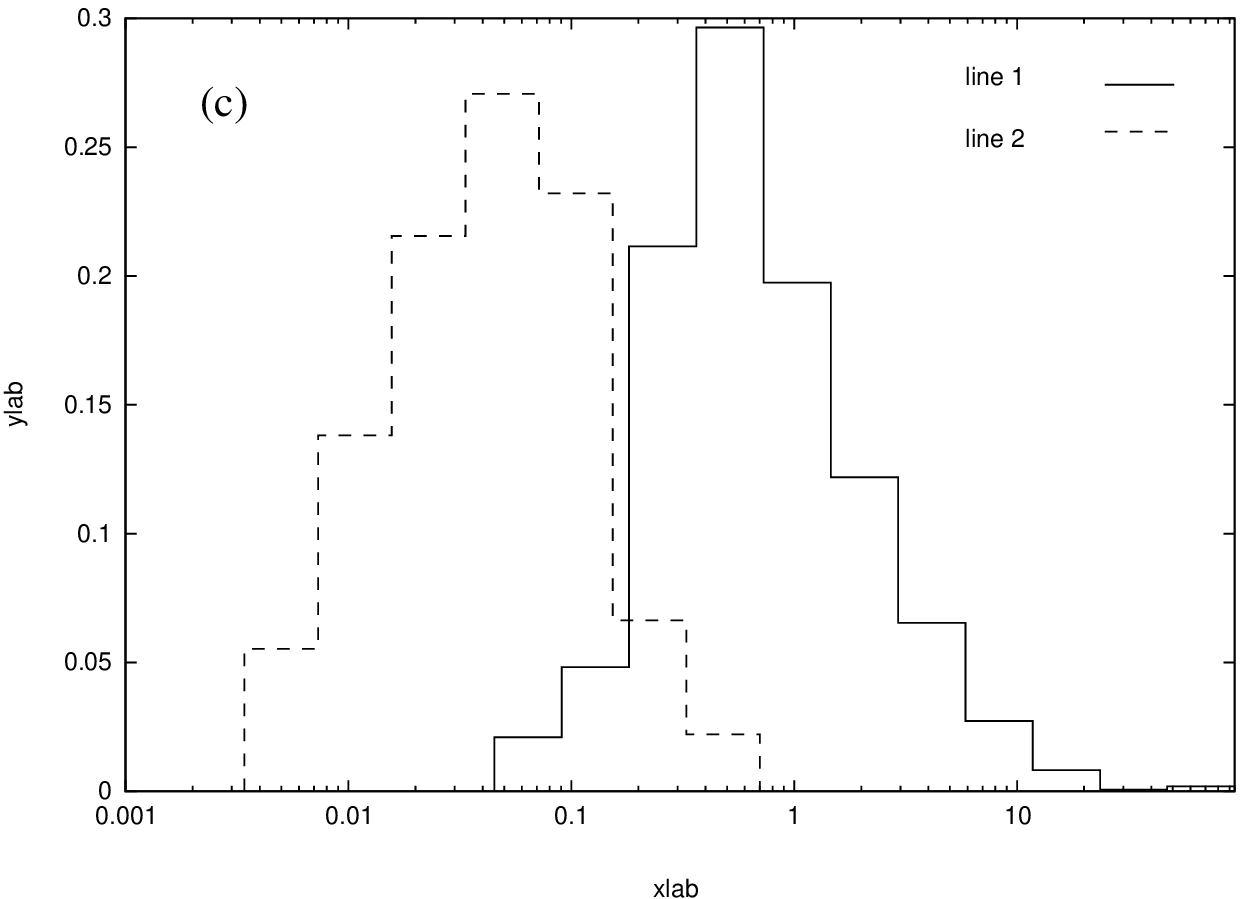} 
\vspace{0.5em}
        \psfrag{xlab}[b]{\small Time Intervals (sec) }
        \psfrag{ylab}[b]{\small Frequency}
\includegraphics[height=5cm]{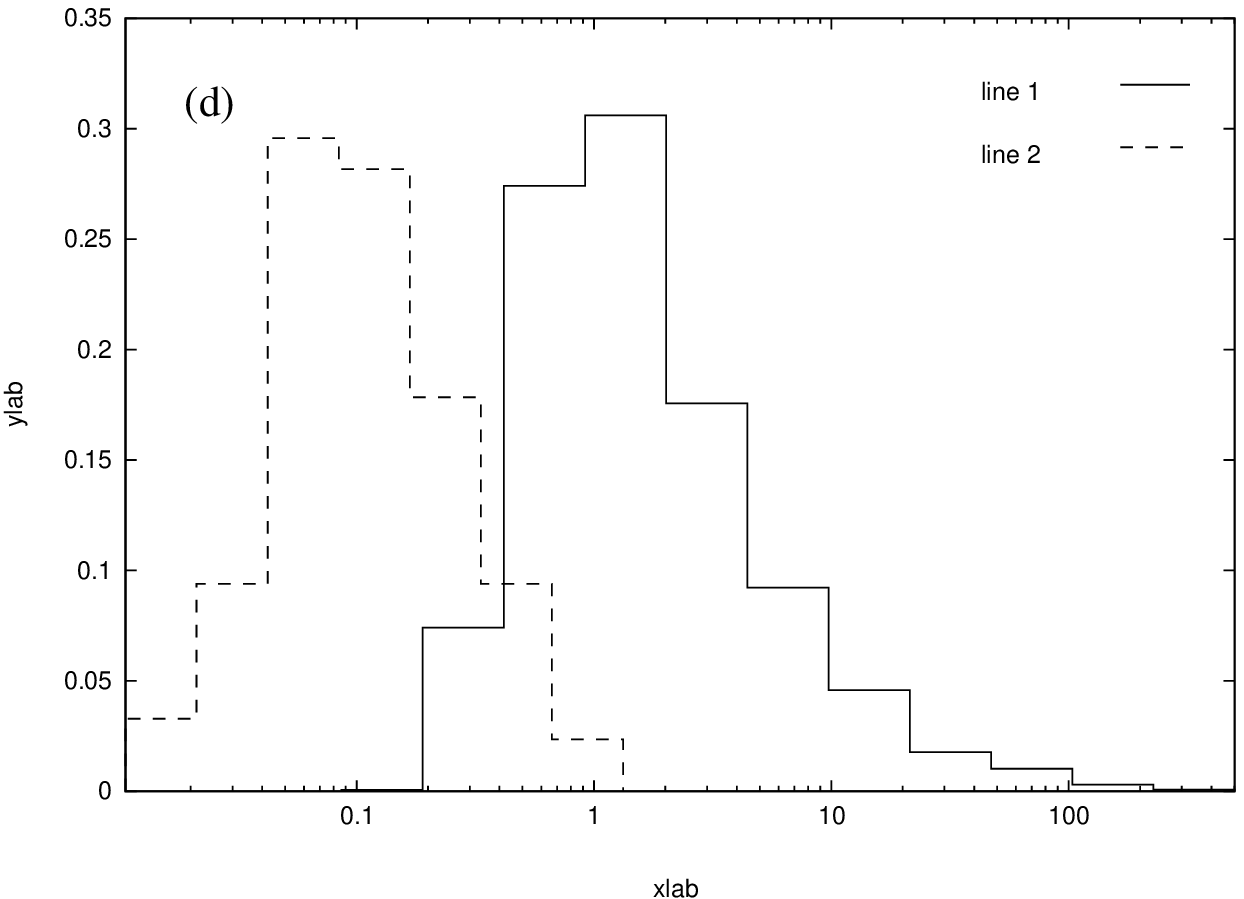} 
\vspace{.5em}
    \caption{
    Normalised distributions of $t_{\rm r}$, $t_{\rm f}$, FWHM, and $\Delta$T (a-d).
      }
\end{center}
\end{minipage}
\end{figure}

\begin{figure}[t]
    \begin{minipage}{\columnwidth}
    \begin{center}
    \leavevmode
\vspace{1.0em}
        \psfrag{xlab}[b]{\small Pulse Amplitudes (count rate)}
        \psfrag{xlab}[b]{\small }
        \psfrag{ylab}[b]{\small Frequency}
    \includegraphics[height=4.4cm]{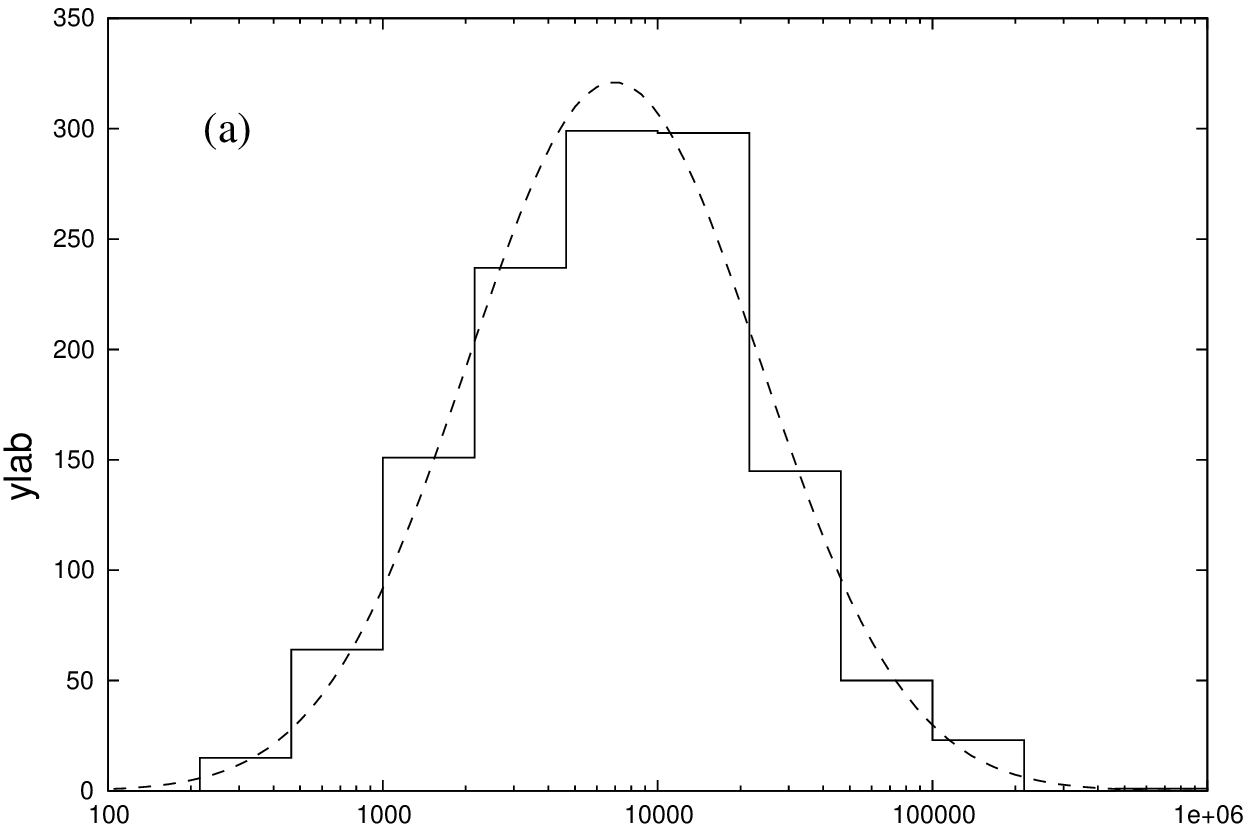}
        \psfrag{xlab}[b]{\small Pulse Amplitudes (count rate)}
        \psfrag{ylab}[b]{\small Frequency}
    \includegraphics[height=4.4cm]{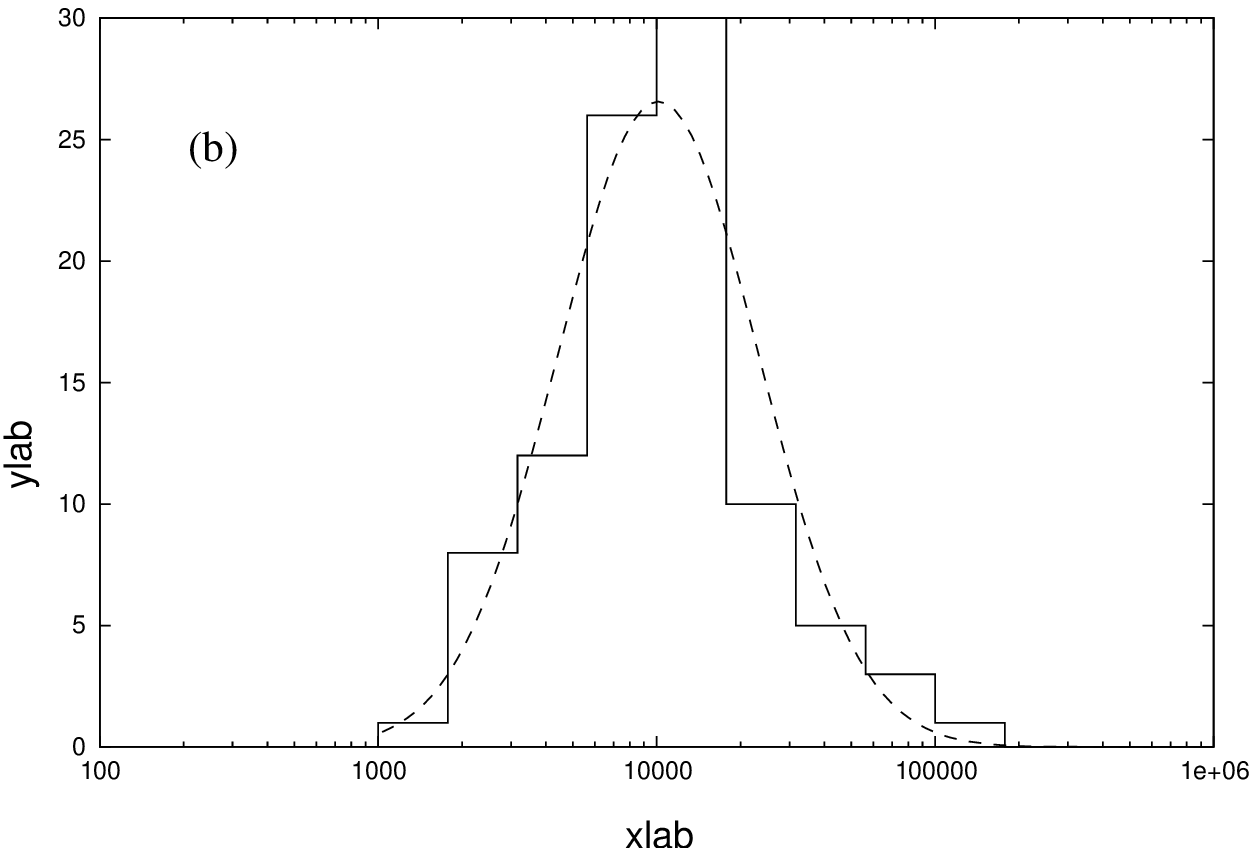}
    \caption
    {Pulse Amplitude distributions for the long GRBs (a) and the short GRBs (b).
    The dashed lines indicate lognormal fits to the data.
    The resolution of the long sample is 64 \,ms and the resolution of the short sample is 5 \,ms.
    }
\end{center}
\end{minipage}
\end{figure}

\begin{figure}[t]
        \begin{minipage}{\columnwidth}
        \begin{center}
    \leavevmode
\vspace{1.0em}
        \psfrag{xlab}[b]{\small Pulse Areas (count rate)}
        \psfrag{xlab}[b]{\small }
        \psfrag{ylab}[b]{\small Frequency}
        \includegraphics[height=4.4cm]{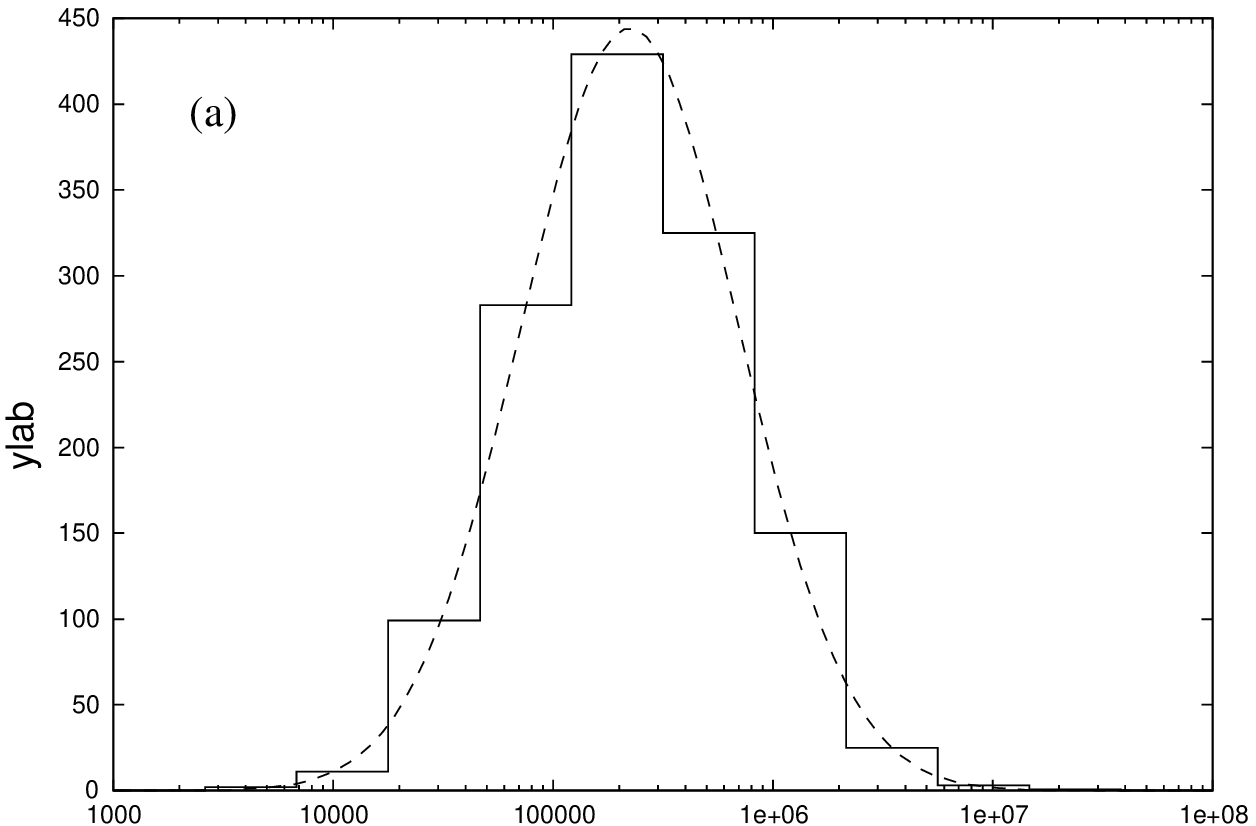}
        \psfrag{xlab}[b]{\small Pulse Areas (count rate)}
        \psfrag{ylab}[b]{\small Frequency}
        \includegraphics[height=4.4cm]{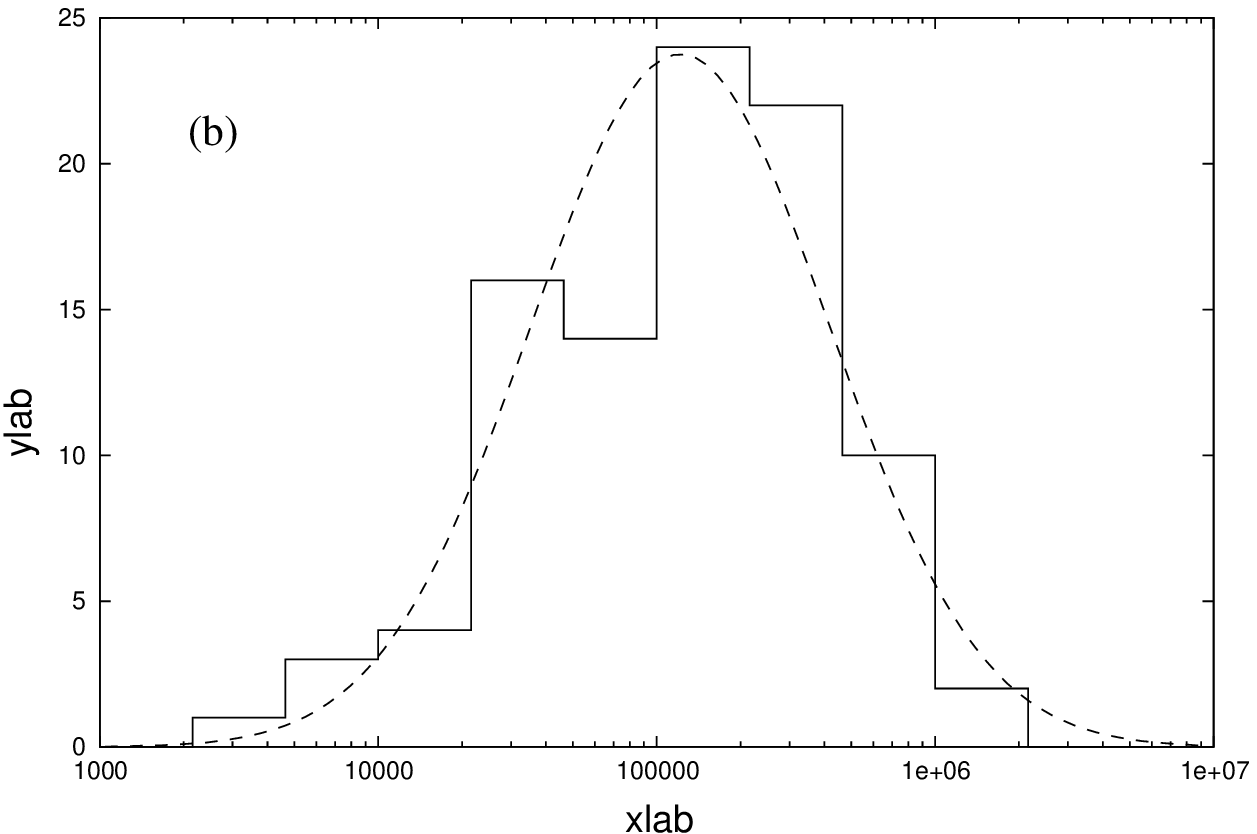}
        \caption
        {Pulse Area distributions for the long GRBs (a) and the short GRBs (b).
    The dashed lines indicate lognormal fits to the data.
    }
\end{center}
\end{minipage}
\end{figure}

\begin{table}[ht]
    \begin{minipage}{\columnwidth}
    \begin{center}
    \begin{tabular}{lcc}   \\
    \hline
    Pulse Property      &   Short GRBs  &   Long GRBs   \\  \hline
    Rise Time (sec)     &   0.035   &   0.64    \\
    Fall Time  (sec)    &   0.056   &   1.10    \\
    FWHM  (sec)         &   0.045   &   0.58    \\
    Time Interval  (sec) &   0.095   &   1.34    \\
    Pulse Area (count rate) &   1.5$\times 10^{5}$  &   1.5$\times 10^{5}$  \\
    Pulse Amplitude (count rate)&   1.0$\times 10^{4}$  &   8$\times 10^{3}$    \\
    \hline
    \end{tabular}
    \caption
    {The median values of the pulse properties in GRBs.
    All pulses with $\mathrm{\tau}_{\rm \sigma}$ $\geq$ 5 were used for the
    time intervals, not just isolated pulses as for the pulse properties.
    }
    \end{center}
\end{minipage}
\end{table}

\begin{table}[h]
    \begin{minipage}{\columnwidth}
    \begin{center}
        \begin{tabular}{lccc}   \\
    \hline
    N   &   Number of Intervals    &      $\rho$ &  Probability    \\ \hline
    1   &   140 &0.42/0.30& $1.7 \times 10^{-12}$/$3.4 \times 10^{-4}$  \\
    2   &   84      &0.48/0.32   & $4.9 \times 10^{-6}$/$3.0 \times 10^{-3}$    \\
    \hline
    1   &   2751    & 0.42/0.56     & $< 10^{-48}$  \\
    2   &   2499    & 0.34/0.48     & $< 10^{-48}$  \\
    5   &   1929    & 0.24/0.37     & $5\times10^{-26}/<10^{-48}$   \\
    10  &   1395    & 0.20/0.29     & $3\times10^{-13}/6\times10^{-27}$ \\
    15 & 890 & 0.16/0.25 & $3\times10^{-6}/4\times10^{-14}$\\
        \hline
        \end{tabular}
        \caption
    {
    Spearman Rank Order correlation coefficients $\rho$ for time intervals between pulses.
    The value of N indicates the number of pulses between the correlated time intervals.
    The two values for $\rho$ and the probability are for unnormalised/normalised time intervals.
    The values are normalised by T$_{\rm 90}$.
    The first two lines refer to the short GRBs and the remaining lines refer to the long GRBs.
    }
    \end{center}
    \end{minipage}
\end{table}

\begin{table}[hbtp]
        \begin{minipage}{\columnwidth}
    \begin{center}
        \begin{tabular}{lccc}   \\
    \hline
        N       &       Number of Pulses &      $\rho$ &  Probability    \\ \hline
    1   &   213/107   &     0.39/0.24       & $5 \times 10^{-9}$/$1.1 \times 10^{-2}$  \\
    2   &   140/84    &     0.13/$-$0.09    & 0.13/0.42  \\
    \hline
    1   &   3039    &   0.72/0.57   &  $< 10^{-48}$\\
    3   &   2499    &   0.55/0.32   &  $< 10^{-48}$\\
    5   &   2098    &   0.52/0.24   &  $< 10^{-48}/ 3\times10^{-29}$\\
    7   &   1777    &   0.48/0.15   & $< 10^{-48}/ 6\times10^{-11}$\\
        \hline
        \end{tabular}
        \caption
    {Spearman Rank Order correlation coefficients $\rho$ for the pulse amplitudes of neighbouring pulses (N=1)
    and pulses separated by N pulses.
    The two values for $\rho$ and the probability are for unnormalised/normalised pulse amplitudes. The later
    are normalised by the maximum pulse amplitude in the burst.
    The first two lines refer to the short GRBs and the remaining lines refer to the long GRBs.
    The maximum peak was removed for the short GRBs in the normalised sample.
    }
    \end{center}
        \end{minipage}
\end{table}

\section{Discussion}
There are remarkable similarities between the statistical
properties of the two sub-classes of GRBs.
The distributions of the $t_{\rm r}$, $t_{\rm f}$, FWHM, pulse amplitude, pulse area and $\Delta$T
for GRBs with T$_{\rm 90}$$<$$2\,$s and  T$_{\rm 90}$$>$$2\,$s are very similar and both are
well described by lognormal distributions \citep{smcb:2001,quillig:2002}.
In long GRBs with T$_{\rm 90}$$>$$2\,$s the values of $\Delta$T are not
random but consistent with a lognormal distribution with a
Pareto-Levy tail for a small number of long time intervals in
excess of 15\,s.
The values of the time intervals between pulses and the pulse amplitudes were found to be correlated
over most of the long GRBs (Tables 2 and 3). In the short GRBs adjacent and subsequent time intervals
 and pulse amplitudes were found to be correlated at a lower significance level due to the smaller number of pulses.

The clear conclusion is
that the same emission mechanism can account for the two types of
GRBs.  This conclusion is in agreement with a very different
analysis of the temporal structure of short GRBs \citep{np:2001}.
The external shock model \citep{derm:1999} has serious difficulties in
accounting for GRBs with T$_{\rm 90}$$<$$2\,$s and with the
non-random distribution of correlated time intervals between pulses.
The results presented here provide considerable support
for the internal shock model \citep{reemes:1994}.  The internal
shock model can account for the results obtained for long and
short GRBs provided the cause of the pulses and the correlated
values of $\Delta$T can be attributed to the central engine.

\section{Conclusions}
Samples of short and long bright GRBs have been denoised
and analysed by an automatic pulse selection algorithm. The
results show that in both cases the distribution of the properties of isolated
pulses and time intervals between all pulses are similar and compatible with
lognormal distributions.  The same mechanism seems to be
responsible for both long and short GRBs and may be attributable to the
internal shock model.


\end{document}